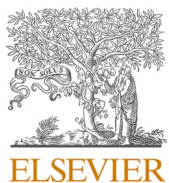

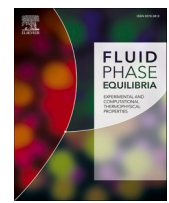

# Exploring pNIPAM lyogels: Experimental study on swelling equilibria in various organic solvents and mixtures, supported by COSMO-RS analysis

Kathrin Marina Eckert [a,*], Simon Müller [a], Gerrit A. Luinstra [b], Irina Smirnova [a]

[a] *Institute of Thermal Separation Processes, Hamburg University of Technology, Hamburg, Germany*
[b] *Institute for Technical and Macromolecular Chemistry, University of Hamburg, Hamburg, Germany*

ABSTRACT

Stimuli-responsive lyogels are considered to be smart materials due to their capability of undergoing significant macroscopic changes in response to external triggers. Due to their versatile and unique properties, smart lyogels exhibit great potential in various applications such as drug delivery or actuation processes. While Poly-N-isopropylacrylamide (pNIPAM) is widely known as a thermo-responsive material, it also shows significant solvent-responsive swelling behavior. For the systems tested, polar solvents induce strong swelling due to hydrogen bonding with the amide group, nonpolar solvents lead to significant shrinkage of the lyogels. The aim of this study is to investigate and to model this behavior for future application in chemical or biochemical reactors. As a current area of research, incorporating smart lyogel technology into (bio)chemical reactors facilitates the development of smart reactor systems. Applying thermodynamic modelling with the g$^E$ model COSMO-RS on the monomer or oligomers of pNIPAM, the correlation between solvent-polymer interactions and the degree of swelling can be observed. pNIPAM derivatives exhibit low infinite dilution activity coefficients (IDACs) in polar solvents with large degrees of swelling, while displaying an increase of IDACs in nonpolar solvents. Hydrogen bonds dominate the swelling behavior of lyogels not only in pure solvents but also in mixtures of solvents with varying polarity. Even in mixtures containing high amounts of nonpolar solvents, large degrees of swelling were observed due to the uptake of the polar solvent in the lyogel matrix. This effect can be observed in binary solvent mixtures but also in representative mixtures along an esterification reaction with varying carboxylic chain length of alcohol and carboxylic acids.

## 1. Introduction

Stimuli-responsive lyogels are capable of undergoing significant macroscopic changes in response to external stimuli such as temperature, pH or solvent composition of the medium [1–5]. The polymer networks have the ability to hold an amount of solvent several times of their own mass without dissolving, because of their characteristic cross-linked three-dimensional structure [6,7]. The uptake or expulsion of solvents caused by changes in the thermodynamic equilibrium results in the macroscopic swelling or shrinking of the lyogel. This unique behavior offers a wide range of possible applications such as responsive actuators [8], drug delivery [9,10], flow control in the form of valves [11,12], or catalysists [13].

Poly-N-isopropylacrylamide (pNIPAM) as thermosensitive polymer [2,14,15] is composed of hydrophilic and hydrophobic functional groups, resulting in amphiphilic behavior and "switchable" properties of the lyogel [16]. The polymer exhibits a *lower critical solution temperature* (LCST) in water, characterized by the formation of hydrogen bonds between polymer chains and water molecules at lower temperatures. Above 307 K, the polymer chains undergo a coil-to-globule transition to a polymer affine state [16]. This transition observed in free polymer chains can be extended to the behavior of three-dimensional lyogels. Below the *volume phase transition temperature* (VPTT), lyogels exhibit strong swelling behavior, whereas increasing the temperature above the VPTT induces lyogel shrinkage [2,16]. In literature, one specific application of pNIPAM lyogels is the creation of a smart reactor through a self-regulating outlet control system [17]. In this process, an exothermic emulsion polymerization was carried out with the polymer acting as a smart valve, switching between the product and waste state. The heat generated during the polymerization reaction was used as a trigger for the thermosensitive lyogel to either swell or shrink. As this reactor set-up relies on the temperature-induced response of lyogels, the conversion rate of a chemical reaction can also serve as a trigger for the lyogel. Hence, exploring the thermodynamic behavior of lyogels in various solvents and mixtures becomes necessary.

Measurements of thermodynamic properties can be performed via various methods adjusted to the system under analysis. Dohrn and Pfohl [18,19] provide general overviews and also polymer-specific experimental methods. While temperature-induced swelling has been extensively studied in both experimental [2,8,20–23] and theoretical reserach [24–26], solvent-induced swelling of lyogels in pure solvents or mixtures are rarely analyzed. Existing literature predominantly focuses on the swelling equilibria of lyogels in water or aqueous solutions, as in Mukae et al. [20,21], Knörgen et al. [22], Scherzinger et al. [23] and Alenichev et al. [27]. These studies on lyogels in aqueous mixtures encompass solvents such as alcohols, tetrahydrofuran, dimethylsulfoxide or acetonitrile [20,24]. Water-free studies focused on lyogels in pure organic solvents and examined the influence of the chosen solvent on the swelling behavior of the lyogels as found in the experimental analysis of Komarova et al. [28], or Martinez et al. [7]. In these publications, pNIPAM primarily exhibits swelling in polar solvents, while nonpolar solvents induce shrinkage behavior [7,28]. The hydrogen bonding between the solvent and polymer leads to solvent uptake in the gel matrix, resulting in larger degrees of swelling [7].

In the literature, the studies on the swelling of lyogels mainly focus on single or binary solvents. Multi-component systems have been primarily studied for pharmaceutical applications. These studies contain aqueous mixtures, for instance, to analyze the impact of media for bacteria growths [29], or for application in tissue engineering potentially as a drug-delivery system [30] using temperature or pH as an external trigger. Redirecting the attention to the application of stimuli-responsive lyogels in chemical or biochemical reactors, previous studies have explored swelling behavior in the presence of chemical or biochemical reactions. These investigations have specifically analyzed swelling induced by stimuli such as temperature [17,31], pH [32], concentration of a specific component or catalyst[33], or chemical reactions involving the lyogel backbone [34–36].

Various theoretical studies of swelling equilibria have been conducted, predominantly centered on Florys theory and its modifications, involving adjustments of the Flory-Huggins parameter to different polymers [37–41]. Computationally more expensive methods, such as molecular dynamics simulations, achieve greater interpretability [26, 42] and lead to deeper insights into polymer-solvent interactions. As alternative approach, thermodynamic modelling on the basis of modified $g^E$ models or equations of state that account for the elastic network behavior of lyogels, has also been performed [24,25,43]. These approaches provide deeper thermodynamic insights into the fluids and the lyogel network while being computationally less expensive than molecular dynamics simulations. For a fully predictive modeling approach in this work, the quantum-chemistry-based $g^E$ model COSMO-RS [44–46] (COnductor like Screening MOdel for Real Solvents) was employed.

In previous studies, COSMO-RS calculations were successfully applied to analyze vapor-liquid equilibria [47], or calculate infinite dilution activity coefficients to perform solubility studies [48] of polymer chains in varying solvents. In these studies, the combinatorial (nonideal entropic) contribution was adjusted, to accurately represent the polymer properties [47,48]. However, these adjustments primarily address the behavior of polymer chains rather than three-dimensional lyogels. Considering the limited configuration changes that the polymeric chains of a lyogel network can undergo, common approaches for polymer calculations [48,49] do not sufficiently describe the behavior of a lyogel network. Nevertheless, this study utilizes the predictive COSMO-RS model, but focuses on the behavior of oligomers in various solvents rather than large polymer structures. Furthermore, it represents the first application of the COSMO-RS model to analyze the relationship between calculated thermodynamic properties and experimentally obtained degrees of swelling of lyogels instead of non-cross-linked polymer chains.

This work combines experimental and theoretical analysis, investigating the influence of different solvents and solvent mixtures (in chemical reactions) on the thermodynamic properties of the lyogels. A particular focus of this study is to examine the impact of functional groups with varying carbon chain lengths on swelling behavior of lyogels for potential applications in chemical or biochemical reactors. To achieve this objective, the study begins by analyzing swelling behavior in pure organic solvents and binary solvent mixtures, exploring polymer-solvent interactions. These findings are subsequently extended to multi-component reactive systems, evaluating the potential application of lyogels in (bio)chemical processes in the future.

## 2. Experimental

### 2.1. Materials

The following chemicals and organic solvents were used without any further additional preparation: N-isopropylacrylamide (NIPAM) (> 98 %), N,N'-methylenebis(acrylamide) (MBA) (> 99 %), ammonium persulfate (> 98 %), and sodium persulfate (> 97 %), acetic acid (≥ 99.8 %), 1-butanol (≥ 99.5 %), butyric acid (99.0 %), butyl acetate (≥ 99.5 %), butyl butyrate (≥ 98.0 %), butyl propionate (≥ 98.0 %), ethanol (≥ 99.9 %), ethyl acetate (≥ 99.7 %), ethyl butyrate (> 99.0 %), ethyl propionate (≥ 99.0 %), methanol (≥ 99.8 %), methyl acetate (≥ 99.0 %), methyl butyrate (≥ 98.0 %), methyl propionate (> 99.0 %), 1-propanol (≥ 99.5 %), propionic acid (≥ 99.0 %), propyl acetate (≥ 98.0 %), propyl butyrate (≥ 98.0 %), propyl propionate (> 99.0 %), valeric acid (≥ 99.0 %). (Table S1)

### 2.2. Gel formulation

For the preparation of hydrogels, 2.175 g NIPAM and 0.030 g MBA were dissolved in 21 g of deionized water, and degassed with nitrogen for 30 min. Subsequently, 2.5 mg of sodium persulfate and ammonium persulfate were each dissolved in 1 mL of deionized water and added to the mixture. The hydrogel monoliths were cast in syringes (BRAUN, 10 mL) and covered with PARAFILM (Sigma-Aldrich).

### 2.3. Swelling equilibria of Lyogels in organic solvents and mixtures

The hydrogel was sliced into cylindrical segments measuring 1–1.5 cm in height and 1.5 cm in diameter; inhomogeneous end pieces were discarded. For the solvent exchange from water to pure solvents or mixtures, the gel monoliths were immersed in the desired solvent or mixture and shaken, followed by two additional solvent exchange steps with fresh solvents each after 24 h. Subsequently, the lyogel was equilibrated for 48 h at 25 °C (± 0.1 °C) in a water bath (Grant OLS200). The required timeframe was determined based on previous experiments. The solvent exchange to pure esters or 1-butanol required an additional solvent exchange to ethanol due to the limited miscibility of the solvents with water. In solvent mixtures, the hydrogel was first immersed in the solvent in which lower swelling behavior of the gel was obtained. Afterwards, the solvent exchange to the requested solvent mixture was performed. The equilibrated lyogels were dried at 40 °C (± 0.1 °C) under vacuum conditions for 48 h to calculate the mass-specific degree of swelling: $Dos = \frac{m_{equilibrated-gel}}{m_{dried-polymer}}$. For the investigation of the swelling behavior during an ongoing esterification reaction, solutions of equimolar mixtures of the reactants were prepared with different concentrations of esters. The concentrations of esters in these mixtures were set to be equivalent to specific conversion rates of the reactions (Fig. 1). In this work, solvent-polymer interactions were investigated by preparing solvent mixtures based on conversion rates of the esterification reaction, without carrying out the reaction experimentally. This method enabled the investigation of the influence of solvent-polymer interactions in the occurring solvent mixtures during the reaction, separate from the reaction kinetics. The solvent mixtures were prepared water-free to avoid the formation of a second solvent phase.

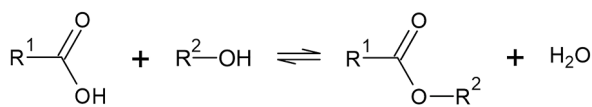

**Fig. 1.** Reaction equation of an esterification reaction. R1 and R2 each describe variable chain lengths of carboxylic acid and alcohol, resulting in variating carbon chain length of the resulting ester.

The composition of the solvent phase after the equilibration of the gel was analyzed by GC (GC 7890 B, Agilent Technologies). The GC method can be described as follows: (*i*) column: DB-WAX, Agilent Technologies (30 m x 0.25 mm x 0.5 μm); (*ii*) column oven temperature: 353.2 °C for 2 min; (*iii*) temperature ramps: 413.2 K (rate = 10 K/min) for 2 min, 483.2 K (rate = 20 K/min) for 5.5 min, (*iv*) carrier gas: nitrogen; (*v*) flow rate: 25 mL/min; (*vi*) injector temperature: 523.2 K; and (*vii*) detector temperature: 533.2 K. Injection was done with a split ratio of 400 and the injection volume was 1 μL.

## 3. Computational methods

### 3.1. Calculation of infinite dilution activity coefficients

The COSMO-RS calculations were performed using the software COSMO*therm* [44–46,50] (BP_TZVPD_FINE_19). For the conformer search COSMO*conf* [44–46,51] (v3.0) with Turbomole [52] (v. 6.6) was used. In this work, different monomer and oligomer structures (trimer, pentamer, heptamer) were investigated and their infinite dilution activity coefficients (IDAC) in various organic solvents $\ln(\gamma_i^\infty)$ were calculated using COSMO*therm* according to Putnam et al. [53].

## 4. Results and discussion

### 4.1. Influence of functional groups and carbon chain length on swelling behavior

This study explores the interactions of pNIPAM in diverse solvents, with a specific focus on the impact of various functional groups. To achieve this, a comparative analysis of carboxylic acids, alcohols, and esters is conducted. Additionally, we systematically vary the carbon chain length of all compounds to investigate its influence on the swelling behavior and overall occurring interactions in pure organic solvents, binary mixtures, and reactive mixtures.

In the initial series of experiments, the response of lyogels when immersed in pure solvents were examined. The swelling or shrinkage behavior of the lyogels is influenced by the interactions between the functional groups of the solvents and those of the polymer. This investigation aimed to unveil the relationship of these interactions with the lyogels' swelling response. The amphiphilic structure of the polymer enables strong hydrogen bonds with the polar amide group, alongside with weaker van der Waals forces involving the polymer backbone and the isopropyl groups. Consequently, it is expected that the swelling behavior of the lyogels is affected by the functional groups and the length of the carbon chains in the solvents. The obtained results, depicted in Fig. 2, agree with this rationale.

Fig. 2 and Fig. 3 compare the degree of swelling of pNIPAM in carboxylic acids, alcohols, and esters derived from alcohols or carboxylic acids with varying carbon chain lengths. Thereby, pNIPAM exhibits a greater degree of swelling in carboxylic acids compared to alcohols, while the lowest degree of swelling is observed in esters. When exploring the distinct solvent groups, it becomes evident that the largest swelling occurs in highly polar protic solvents, while less polar solvents induce shrinkage. These observations align with existing literature findings [7,15,28] and result from the specific interactions between solvents and the polymer. Carboxylic acids and alcohols primarily form hydrogen bonds with the polymer, whereas esters are not expected to show significant hydrogen bond formation. The increase in carbon chain length of esters reduces solvent polarity, weakening interactions with the polymer and resulting in decreased swelling. In contrast, the hydrogen bonds of carboxylic acids or alcohols with pNIPAM remain unaffected by an increasing carbon chain, probably due to the higher strength of these interactions. For these solvent classes, the degree of swelling slightly increases with the growing carbon chain length. It can be assumed, that hydrogen bonds stabilize solvent molecules within the polymer structure. This leads to an increase in swelling degrees with longer chain lengths due to the larger sizes of the solvent molecules.

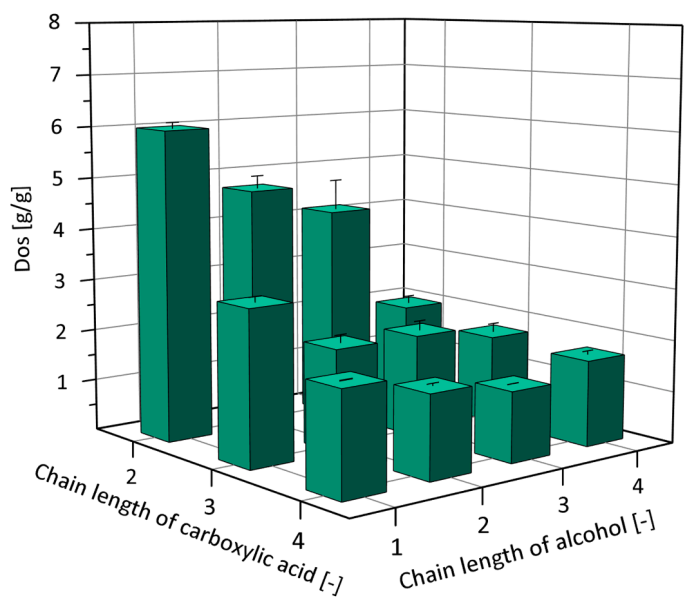

**Fig. 3.** Degree of swelling (Dos) of pNIPAM lyogels in esters with carbon chain lengths. The carbon chain length of the carboxylic acid (acetic acid, propionic acid, butyric acid) and the alcohol (methanol, ethanol, 1-propanol, 1-butanol) from which the ester is formed are varied. The standard deviations are calculated based on multiple experiments (*n* = 3).

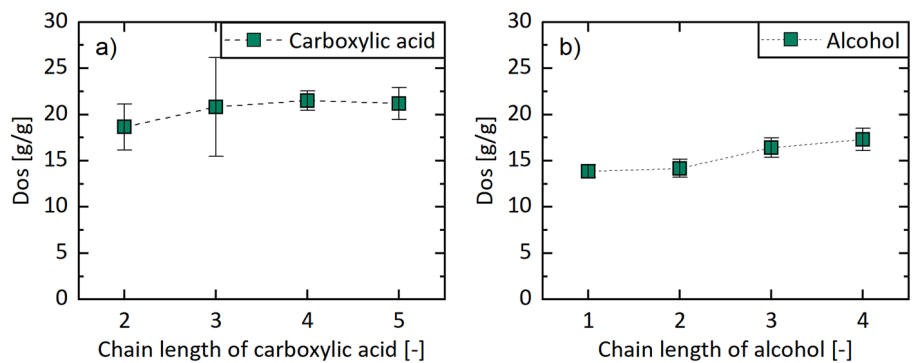

**Fig. 2.** Degree of swelling (Dos) of pNIPAM lyogels in pure organic solvents. Varying carbon chain length of a) carboxylic acids (acetic acid, propionic acid, butyric acid, valeric acid), b) alcohols (methanol, ethanol, 1-propanol, 1-butanol). The dashed lines are added to guide the eye. The standard deviations are calculated based on multiple experiments (*n* = 3).

In addition to primary alcohols, the degree of swelling of pNIPAM in the secondary alcohol 2-propanol was investigated. The obtained degree of swelling of 11.48 ± 0.91 g/g is significantly lower than in the isomer 1-propanol (Table S2). A possible reason for this may be the reduced accessibility of the hydroxyl group, resulting in weaker or fewer hydrogen bonds formed between solvent and polymer.

Further, the swelling behavior of pNIPAM in esters was studied systematically by variation of the carbon chain length of both the alcohol and the carboxylic acid from which the ester is formed (Fig. 3). In esters formed from propionic or butyric acid almost constant degrees of swelling are observed regardless of the carbon number of the alcohol in the ester molecules. On the contrary, altering the carbon number of the carboxylic acid in the ester yields constant swelling degrees only in butyl esters. These findings suggest that carbon chains length of the carboxylic acid has a greater impact on the swelling behavior than carbon chain lengths of the alcohol group in the corresponding ester molecules.

It could be shown that the carbon chain length of both the alcohol and carboxylic acid in ester molecules influences the degree of swelling in these solvents. However, longer chains, as found in esters composed of 1-butanol or butyric acid, dominate the swelling behavior due to their significant impact. Aiming for the application of solvent-triggered actuation in reactive systems, changing swelling degrees with the progression of the chemical reaction is necessary. Concluding from the previous findings, reactions involving polarity changes (comparing reactants and products) are crucial for this type of actuation. Moreover, it is evident that the application of lyogels in reactive systems requires detailed knowledge about the polarity and the interactions of solvents with the polymer, as it significantly influences the swelling behavior.

To enable deeper insights into these interactions, a theoretical study was conducted using the $g^E$ model COSMO-RS, and compared with the results of the previous experimental investigations. The model was selected for its predictive nature as potential screening tool assessing the compatibility of selected polymers with the considered reaction system. In this work, the software COSMO*therm* [44–46,50] (BP_TZVPD_FINE_19) was utilized, but the open-source implementation openCOSMO-RS [54] can also be applied for these purposes. As indicators of intermolecular interactions underlying thermodynamic equilibria, the infinite dilution activity coefficients and interaction energies (hydrogen bond, van der Waals) in mixtures are investigated in the following section. Due to the long calculation times, this analysis focuses not on entire polymers, but on different building blocks of the polymers (3-mers, 5-mers, 7-mers; Table S3) as a faster calculation approach.

A negative IDAC value indicates that the solvents interact stronger with the polymer structure rather than with itself. Based on the experimental investigations, strong interactions are expected to occur in carboxylic acids, followed by alcohols and esters showing weakest polymer-solvent interactions. Consequently, the smallest IDAC values are expected for carboxylic acids, followed by alcohols, and with esters having the largest IDAC values.

The calculated data (Table 1) of the IDACs aligns with the formulated expectations. Among all pNIPAM derivatives (Table S4), carboxylic acids exhibit the smallest IDAC values, followed by alcohols and esters. Considering this, the model emphasizes further that the overall interactions between acids or alcohols and pNIPAM are stronger than those between esters and pNIPAM. Existing literature attributes this phenomenon to the formation of hydrogen bonds between polar protic solvents and pNIPAM [7,28]. This aligns with the calculated hydrogen bond interaction energies, which are strongest in polar protic solvents (carboxylic acids followed by alcohols) and weakest in nonpolar (ester) solvents. However, the van der Waals interactions remained constant in all calculations (Figure S4, Table S7). This indicates that the hydrogen bond interactions dominate the overall swelling behavior of the lyogels.

Since the COSMO-RS calculations align with the expectations from the experimental analysis and reported literature data, the following section explores possible correlations of the experimental results and COSMO-RS data.

In Fig. 4, the experimentally determined degree of swelling is presented as a function of IDACS calculated by the COSMO-RS model. All data series exhibit a similar trend of decreasing swelling degree with increasing IDACs. This trend is most pronounced in oligomer molecules and least apparent in monomers. The obtained trend may be attributed to the progressive approximation of the building blocks to the crosslinked polymer structure in the lyogel. The structure is characterized by fewer double bonds and a higher number of monomeric units resulting from radical polymerization synthesis. Nevertheless, all pNIPAM derivatives show a strong decrease in swelling beyond a certain threshold value for the IDAC.

Considering that activity coefficients describe non-ideal thermodynamic behavior resulting from intermolecular interactions, IDACs are commonly utilized in the literature to characterize solvent-solute interactions [55]. The obtained trend (Fig. 4) underscores the transferability of this method to the swelling behavior of lyogels. It could be shown that the calculated IDACs can be correlated with the swelling behavior of lyogels in different solvents by investigating pNIPAM derivatives.

**Table 1**
IDACs and interaction energies of NIPAM in organic solvents.

| Organic Solvent | IDACs of monomer in organic solvent [-] | Hydrogen bond interaction energy [kJ/mol] | Van der Waals interaction energies [kJ/mol] |
|---|---|---|---|
| Acetic acid | −1.032 | −14.413 | −82.341 |
| Propionic acid | −1.330 | −14.458 | −82.848 |
| Butyric acid | −1.309 | −14.179 | −83.095 |
| Methanol | −0.248 | −11.362 | −80.038 |
| Ethanol | 0.026 | −9.910 | −81.186 |
| 1-Propanol | −0.010 | −9.688 | −81.935 |
| 1-Butanol | 0.057 | −9.349 | −82.358 |
| Methyl acetate | 0.352 | −1.402 | −82.452 |
| Ethyl acetate | 0.410 | −1.545 | −82.710 |
| Propyl acetate | 0.60759 | −1.489 | −82.879 |
| Butyl acetate | 0.813 | −1.369 | −83.003 |
| Methyl propionate | 0.582 | −1.306 | −82.849 |
| Ethyl propionate | 0.732 | −1.361 | −82.964 |
| Propyl propionate | 0.900 | −1.266 | −83.029 |
| Butyl propionate | 1.101 | −1.186 | −83.320 |
| Methyl butyrate | 0.800 | −1.136 | −83.054 |
| Ethyl butyrate | 0.870 | −1.209 | −83.036 |
| Propyl butyrate | 1.173 | −1.181 | −83.135 |
| Butyl butyrate | 1.271 | −1.197 | −83.091 |

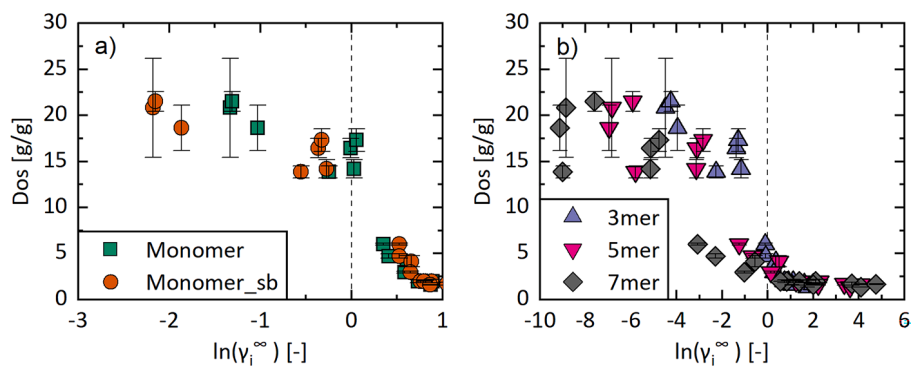

**Fig. 4.** Degree of swelling over infinite dilution activity coefficients (IDACs) for different pNIPAM building blocks and the analyzed solvents. a) monomer (N-(propan-2-yl)prop-2-enamide) (green squares) and monomer without double bond (N-(propan-2-yl)prop-2-anamide) (orange circles), b) trimer (purple triangles), pentamer (pink inverted triangle) and heptamer (gray diamond) of NIPAM. The dashed line highlights $\ln(\gamma_i^\infty) = 0$ where the mixture behaves ideally. The standard deviations are calculated based on multiple experiments ($n = 3$).

The next step is the comparison of the calculated IDACs with the hydrogen bond interaction energies. This allows for a further examination of occurring polymer-solvent interactions, with the results being utilized for a qualitative comparison among the considered solvents.

As shown in Fig. 5, the hydrogen bond interaction energies can be reasonably correlated with the experimentally determined degrees of swelling. It highlights the significant influence of occurring hydrogen bond interaction energies on the swelling of lyogels, as can be also seen in Table 1. Moreover, Fig. 5a) and b) clearly illustrate the correlation between IDACs and hydrogen bond interaction energies: the stronger the occurring hydrogen bonds, the more pronounced is the IDACs' deviation from unity (ideal solution). This relation successfully enables the qualitative comparison of lyogel swelling in various organic solvents in a fully predictive approach.

From the previous investigations we can conclude, that the calculations using COSMO-RS allow to reveal and quantify the mechanisms of interactions determining the swelling behavior of pNIPAM lyogels. The comparisons involving hydrogen bonds and infinite dilution activity coefficients can serve as a powerful tool to predict the swelling behavior of lyogels in different solvents. Unlike other models in the literature [24, 40,41,43], these qualitative comparisons can be applied without the need for fitting parameters. The utilization of the COSMO-RS model enables the rationalization to explain the occurring polymer-solvent interactions. In the context of implementing lyogels in reactive systems, COSMO-RS based investigations can further assist in screening

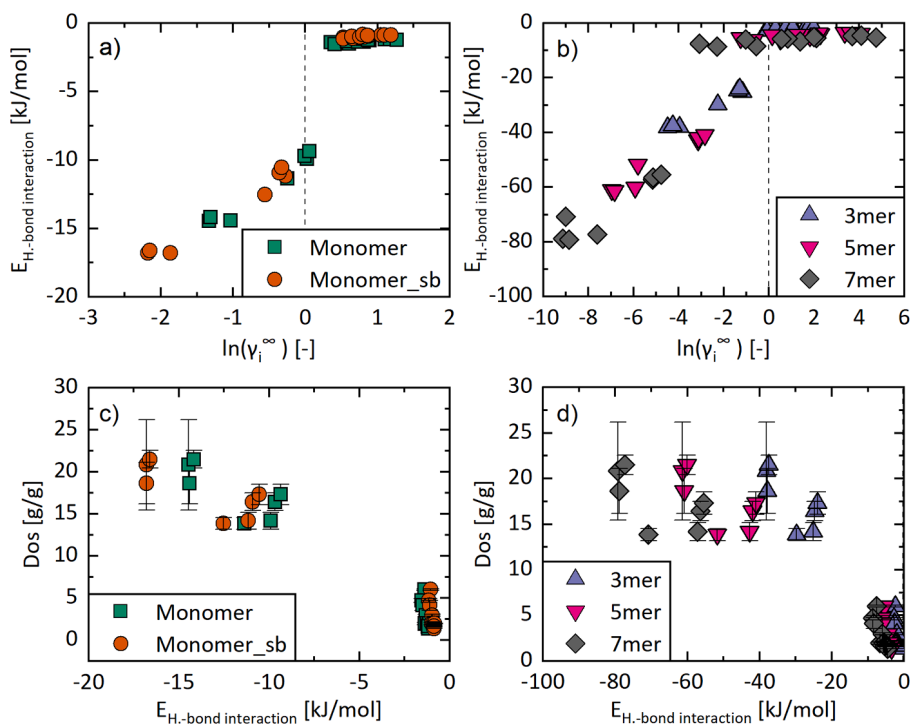

**Fig. 5.** Hydrogen bond interaction energies of infinitely diluted pNIPAM building blocks in pure organic solvents depending on infinite dilution activity coefficients depicted in a) and b), and degrees of swelling in c) and d). The considered building blocks are: monomer (N-(propan-2-yl)prop-2-enamide) (green squares) and monomer without double bond (N-(propan-2-yl)prop-2-anamide) (orange circles), B: trimer (purple triangles), pentamer (pink inverted triangle), heptamer (gray diamond). The dashed line highlights $\ln(\gamma_i^\infty) = 0$ where the mixture behaves ideally. The standard deviations are calculated based on multiple experiments ($n = 3$).

suitable polymers or tailoring specific polymers for specific chemical or biochemical reactions.

## 4.2. Swelling behavior of pNIPAM lyogels in binary solvent mixtures

As a next step, we investigated water-free mixtures of solvents containing diverse functional groups to analyze their impact on the swelling behavior and solvent uptake of pNIPAM. Applying the previously gained insights in polymer-solvent interactions, binary solvent mixtures are investigated, including esters with varying carbon chain lengths and the corresponding alcohols or carboxylic acids. Based on the results of COSMO-RS predictions, mixtures with low ester content are expected to exhibit strong swelling. This swelling is anticipated to decrease with increasing ester contents, attributed to the reduced formation of hydrogen bonds. Fig. 6 shows the degree of swelling as a function of the composition of the bulk liquid.

The overall trend in the swelling behavior of pNIPAM lyogels in binary solvent mixtures remains consistent across all systems. Pure solvents exhibit the extreme (smallest and largest) degrees of swelling for each system: methyl and butyl acetate yielding the smallest swelling and methanol, 1-butanol, and acetic acid resulting in the largest degrees of swelling. Contrary to the expected decrease in swelling with increasing ester content, all systems exhibit large swelling ($> 10$ g/g) in binary mixtures (Fig. 6), regardless of the ester content. This can be attributed to the formation of hydrogen bonds between carboxylic acids and alcohols with pNIPAM lyogels. These interactions lead to a strong uptake of these solvents into the lyogel matrix, leading to larger degrees of swelling even at high ester contents. These observations highlight that, although the theoretical investigations enable the quantification and comprehension of polymer-solvent interactions in pure solvents, they cannot fully predict the behavior of lyogels in mixed solvents.

Fig. 6 further shows that pNIPAM lyogels exhibit similar swelling in carboxylic acid-containing mixtures, with both methyl acetate and butyl acetate. In alcohol-containing mixtures, those with butyl acetate lead to smaller swelling of pNIPAM compared to those with methyl acetate. This phenomenon can be attributed to the stronger interactions of carboxylic acids and pNIPAM, which remain unaffected by the presence of esters. In contrast, interactions between alcohol and pNIPAM are weakened by the presence of esters, particularly for those with a higher nonpolar behavior. These findings align with the results presented in the preceding section on lyogel swelling in pure solvents. It can be concluded that in binary mixtures, the swelling behavior is significantly influenced by the carbon chain lengths of the solvents; however, it is still predominately influenced by the functional groups and the hydrogen bonds formed between the polymer and the solvent mixtures.

The investigations of swelling behavior in pure solvents and in binary solvent mixtures provide crucial insights in occurring polymer-solvent interactions concerning different functional groups and their impact on the degree of swelling. These findings are essential for the potential future application of lyogels in chemical or biochemical reactors, as they determine the suitability of the lyogels to the reaction system and possibly self-regulation of the process. Table 2 gives an overview of the experimental results and their analysis.

## 4.3. Swelling behavior of pNIPAM lyogels in reactive systems

Building upon the insights gained from swelling properties in pure solvents and binary mixtures (Table 2), the final step toward integration of lyogels into chemical or biochemical reactors involves the extension to reactive, multi-component systems. In total, the swelling behavior of pNIPAM lyogels was measured in 12 esterification systems with varying chain lengths of the alcohol and the carboxylic acid. The degrees of swelling were measured at various equivalent ester concentrations, prepared according to specific conversion rates of esterification reactions.

With regard to the behavior of pNIPAM lyogels in the previous systems (Table 2), it can be assumed that large swelling degrees can be obtained even at high equivalent ester concentrations, attributed to the formation of hydrogen bonds between the carboxylic acids and alcohols with the polymer. In presence of polar protic solvents, hydrogen bonds dominate the swelling behavior of pNIPAM. Thus, only small differences between the systems are expected. Furthermore, the lowest degrees of swelling are expected to occur in pure esters, equivalent to 100 % conversion.

**Table 2**
Overview of investigations about swelling behavior of pNIPAM in pure solvents and binary solvent mixtures.

| Solvent type | pNIPAM in pure solvents | pNIPAM in binary mixtures of esters with alcohols / carboxylic acids |
|---|---|---|
| Carboxylic acid | Large Dos ($> 15$ g/g), increases with increasing carbon chain length (Fig. 2); Strong hydrogen bond interaction energies, small IDAC values (COSMO-RS) (Fig. 4, Fig. 5); | Large Dos ($> 15$ g/g), no influence of ester on Dos (Fig. 6); Hydrogen bonds not effected by ester presence (Fig. 6); |
| Alcohol | Large Dos ($> 10$ g/g); increases with increasing carbon chain length (Fig. 2); Strong hydrogen bond interaction energies, small IDAC values (COSMO-RS) (Fig. 4, Fig. 5); | Large Dos ($> 10$ g/g), slightly decreased in ester with higher hydrophobicity (Fig. 6); Hydrogen bonds influenced by nonpolar solvents (Fig. 6); |
| Ester | Small Dos ($< 10$ g/g), decreases for increasing carbon chain lengths (Fig. 3); Negligible hydrogen bond interaction energies, large IDAC values, indicating less favorable solvent (COSMO-RS) (Fig. 4, Fig. 5); | Large Dos ($> 10$ g/g), influence of ester depends on polymer-solvent interactions in mixture (Fig. 6); Hydrophobicity of ester determines impact on Dos (Fig. 6); |

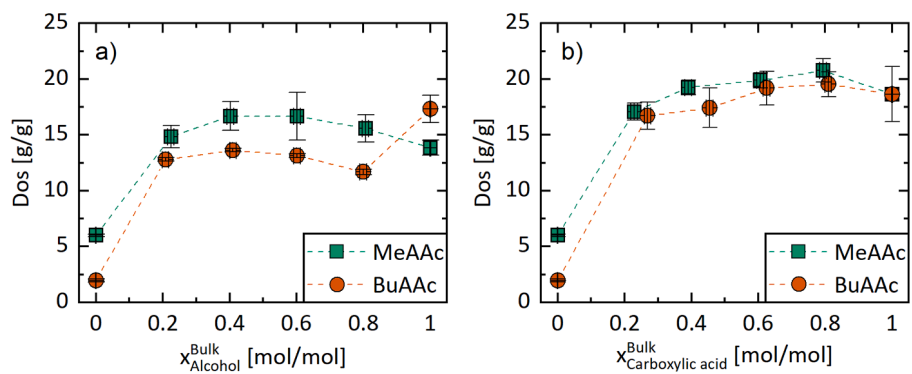

**Fig. 6.** Degree of swelling of pNIPAM lyogels in binary solvent mixtures containing carboxylic acids, or alcohols with esters. On the x-axis, the mole fraction of the carboxylic acid or alcohol in the bulk mixture is plotted. The considered mixtures contain: a) methyl acetate with methanol (green squares), or butyl acetate with 1-butanol (orange circles), and b) methyl acetate with acetic acid (green squares), or butyl acetate with acetic acid (orange circles). The dashed lines are added to guide the eye. The standard deviations are calculated based on multiple experiments ($n = 3$).

The swelling behavior in reactive systems (Fig. 7, Figure S5) aligns well with the expectations derived from the findings in the previous sections. Large swelling behavior can be obtained even at higher equivalent ester concentrations, showing only minor influence of the present ester on the swelling behavior.

In mixtures of esters derived from methanol and ethanol, the degree of swelling increases at higher ester concentrations, equivalent to 50–75 % conversion. In comparison, constant degrees of swelling can be observed in mixtures of esters formed from 1-propanol and 1-butanol. This behavior is most significant in acetate systems and least in the butyrate systems. This effect agrees with the previous findings concerning the swelling behavior in binary solvent mixtures of alcohol or carboxylic acid with esters. It highlights that the formation of hydrogen bonds between the polymer and carboxylic acid is not affected by the presence of esters, whereas the hydrogen bonds between the polymer and alcohol are influenced by this fact

Overall, the insights into polymer-solvent interactions allow to predict the behavior of lyogels in the multicomponent reactive systems to certain extent. In future, however, not only the thermodynamic equilibrium properties like final swelling, but also the kinetics of the swelling behavior should be determined to provide a comprehensive understanding of lyogels suitability for the final application in reactor systems.

## 5. Conclusions

This study demonstrated the promising properties of pNIPAM lyogels for solvent-induced swelling in chemical or biochemical processes. We explored the swelling behavior of pNIPAM lyogels in pure organic solvents, binary, and multicomponent reactive mixtures, analyzing the impact of interactions and carbon chain lengths. The swelling behavior in all examined systems is influenced by the functional groups, polymer-solvent interactions, and the carbon chain lengths of the compounds. Based on this, we hypothesized that hydrogen bonds predominantly influence the swelling behavior. The hypothesis was successfully confirmed through COSMO-RS calculations. Utilizing the COSMO-RS model, we have suggested a fully predictive theoretical tool to quantify and compare lyogel swelling behavior based on infinite dilution activity coefficients and interaction energies.

Concurrently, thermodynamic modelling approaches can predictively assist in selecting and evaluating the suitability of polymers for the required processes.

## CRediT authorship contribution statement

**Kathrin Marina Eckert:** Writing – review & editing, Writing – original draft, Visualization, Validation, Supervision, Software, Resources, Methodology, Investigation, Funding acquisition, Formal analysis, Data curation, Conceptualization. **Simon Müller:** Writing – review & editing, Supervision, Methodology, Investigation, Funding acquisition, Formal analysis, Conceptualization. **Gerrit A. Luinstra:** Project administration, Funding acquisition, Conceptualization. **Irina Smirnova:** Writing – review & editing, Writing – original draft, Supervision, Project administration, Funding acquisition, Formal analysis, Conceptualization.

## Declaration of competing interest

The authors declare that they have no known competing financial interests or personal relationships that could have appeared to influence the work reported in this paper.

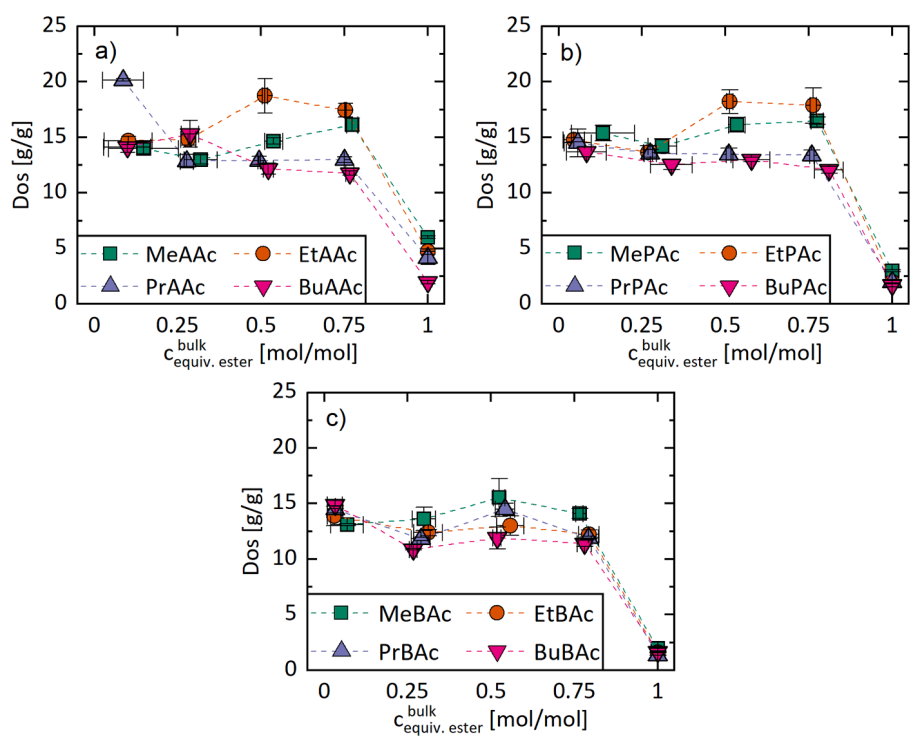

**Fig. 7.** Degrees of swelling of pNIPAM lyogels in ternary solvent mixtures. The bulk phase is prepared with ester concentrations in equimolar mixtures of carboxylic acid and alcohol equivalent to the concentration at certain conversion rates of the esterification reaction. Chain lengths of alcohol and carboxylic acid are variated: a) acetate esters (-AAc), b) propionate esters (-PAc), c) butyric esters (-BAc). Differentiation of the systems regarding the used alcohols: methanol (Me-) (green squares), ethanol (Et-) (orange circles), 1-propanol (Pr-) (purple triangles), 1-butanol (Bu-) (pink inverted triangles). The dashed lines are added to guide the eye. The standard deviations are calculated based on multiple experiments ($n = 3$).

## Data availability

Data will be made available on request.

## Acknowledgment

This project is funded by the Deutsche Forschungsgemeinschaft (DFG, German Research Foundation) – SFB 1615 – 503850735. We especially thank Jana Katharina Ruhstrat and Robin Stumpenhagen for their assistance in the practical experiments presented in this work.

## Supplementary materials

Supplementary material associated with this article can be found, in the online version, at doi:10.1016/j.fluid.2024.114182.

## References

[1] F. Ullah, M.B.H. Othman, F. Javed, Z. Ahmad, H.Md Akil, Classification, processing and application of hydrogels: a review, Mater. Sci. Eng. C 57 (2015) 414–433, https://doi.org/10.1016/j.msec.2015.07.053.

[2] H. Kojima, F. Tanaka, C. Scherzinger, W. Richtering, Temperature dependent phase behavior of PNIPAM microgels in mixed water/methanol solvents, J. Polym. Sci. Part B Polym. Phys. 51 (14) (2013) 1100–1111, https://doi.org/10.1002/polb.23194.

[3] F. Taktak, Rapid deswelling of PDMAEMA hydrogel in response to ph and temperature changes and its application in controlled drug delivery, Afyon Kocatepe Univ. J. Sci. Eng. 16 (1) (2016) 68–75, https://doi.org/10.5578/fmbd.26306.

[4] M.C. Koetting, J.T. Peters, S.D. Steichen, N.A. Peppas, Stimulus-responsive hydrogels: theory, modern advances, and applications, Mater. Sci. Eng. R Rep. 93 (2015) 1–49, https://doi.org/10.1016/j.mser.2015.04.001.

[5] A. Richter, S. Klatt, G. Paschew, C. Klenke, Micropumps operated by swelling and shrinking of temperature-sensitive hydrogels, Lab. Chip 9 (4) (2009) 613–618, https://doi.org/10.1039/B810256B.

[6] Z. Liu, J. Wei, Y. Faraj, X.-J. Ju, R. Xie, W. Wang, L.-Y. Chu, Smart Hydrogels: network design and emerging applications, Can. J. Chem. Eng. 96 (10) (2018) 2100–2114, https://doi.org/10.1002/cjce.23328.

[7] M.V. Martinez, M.A. Molina, S.B. Barbero, Large swelling capacities of crosslinked poly(N-Isopropylacrylamide) gels in organic solvents, MRS. Adv. 3 (63) (2018) 3735–3740, https://doi.org/10.1557/adv.2018.594.

[8] Hydrogel Sensors and Actuators, in: G. Gerlach, K.-F. Arndt (Eds.), Springer Series On Chemical Sensors and Biosensors, Springer, Heidelberg ; New York, 2009. ISBN: 978-3-540-75644-6.

[9] Z. Ayar, M. Shafieian, N. Mahmoodi, O. Sabzevari, Z. Hassannejad, A rechargeable drug delivery system based on pNIPAM hydrogel for the local release of curcumin, J. Appl. Polym. Sci. 138 (40) (2021) 51167, https://doi.org/10.1002/app.51167.

[10] A. Gandhi, A. Paul, S.O. Sen, K.K. Sen, Studies on Thermoresponsive polymers: phase behaviour, drug delivery and biomedical applications, Asian J. Pharm. Sci. 10 (2) (2015) 99–107, https://doi.org/10.1016/j.ajps.2014.08.010.

[11] A. Richter, D. Kuckling, S. Howitz, T. Gehring, K. Arndt, Electronically controllable microvalves based on smart hydrogels: magnitudes and potential applications, J. Microelectromech. Syst. 12 (5) (2003) 748–753, https://doi.org/10.1109/JMEMS.2003.817898.

[12] J. Wang, Z. Chen, M. Mauk, K.-S. Hong, M. Li, S. Yang, H.H. Bau, Self-actuated, thermo-responsive hydrogel valves for lab on a chip, Biomed. Microdevices 7 (4) (2005) 313–322, https://doi.org/10.1007/s10544-005-6073-z.

[13] A. Döring, W. Birnbaum, D. Kuckling, Responsive hydrogels – structurally and dimensionally optimized smart frameworks for applications in catalysis, micro-system technology and material science, Chem. Soc. Rev. 42 (17) (2013) 7391–7420, https://doi.org/10.1039/C3CS60031A.

[14] J. Harrer, M. Rey, S. Ciarella, H. Löwen, L.M.C. Janssen, N. Vogel, Stimuli-responsive behavior of PNiPAm microgels under interfacial confinement, Langmuir. 35 (32) (2019) 10512–10521, https://doi.org/10.1021/acs.langmuir.9b01208.

[15] T. Itahara, T. Tsuchida, M. Morimoto, Solvent-driven swelling and shrinking of poly(NIPAM) gels crosslinked by tris-Methacrylated phloroglucinol derivatives, Polym. Chem. 1 (7) (2010) 1062–1066, https://doi.org/10.1039/C0PY00068J.

[16] J. Šťastná, L. Hanyková, Z. Sedláková, H. Valentová, J. Spěváček, Temperature-induced phase transition in hydrogels of interpenetrating networks poly(N-Isopropylmethacrylamide)/Poly(N-Isopropylacrylamide), Colloid Polym. Sci. 291 (10) (2013) 2409–2417, https://doi.org/10.1007/s00396-013-2992-z.

[17] X. Hu, J. Karnetzke, M. Fassbender, S. Drücker, S. Bettermann, B. Schroeter, W. Pauer, H.-U. Moritz, B. Fiedler, G. Luinstra, I. Smirnova, Smart reactors — Combining stimuli-responsive hydrogels and 3D printing, Chem. Eng. J. 387 (2020) 123413, https://doi.org/10.1016/j.cej.2019.123413.

[18] O. Pfohl, R. Dohrn, Provision of thermodynamic properties of polymer systems for industrial applications, Fluid. Phase Equilib. 217 (2) (2004) 189–199, https://doi.org/10.1016/j.fluid.2003.06.001.

[19] R. Dohrn, O. Pfohl, Thermophysical properties—Industrial directions, Fluid. Phase Equilib. 194–197 (2002) 15–29, https://doi.org/10.1016/S0378-3812(01)00791-9.

[20] K. Mukae, M. Sakurai, S. Sawamura, K. Makino, S.W. Kim, I. Ueda, K. Shirahama, Swelling of poly(N-Isopropylacrylamide) gels in water-aprotic solvent mixtures, Colloid Polym. Sci. 272 (6) (1994) 655–663, https://doi.org/10.1007/BF00659279.

[21] K. Mukae, M. Sakurai, S. Sawamura, K. Makino, S.W. Kim, I. Ueda, K. Shirahama, Swelling of poly(N-Isopropylacrylamide) gels in water-alcohol (C1-C4) mixed solvents, J. Phys. Chem. 97 (3) (1993) 737–741, https://doi.org/10.1021/j100105a034.

[22] M. Knörgen, K.-F. Arndt, S. Richter, D. Kuckling, H. Schneider, Investigation of swelling and diffusion in polymers by 1H NMR imaging: LCP networks and hydrogels, J. Mol. Struct. 554 (1) (2000) 69–79, https://doi.org/10.1016/S0022-2860(00)00560-3.

[23] C. Scherzinger, A. Schwarz, A. Bardow, K. Leonhard, W. Richtering, Cononsolvency of Poly-N-Isopropyl Acrylamide (PNIPAM): microgels versus linear chains and Macrogels, Curr. Opin. Colloid Interface Sci. 19 (2) (2014) 84–94, https://doi.org/10.1016/j.cocis.2014.03.011.

[24] M.C. Arndt, G. Sadowski, Modeling poly(N -Isopropylacrylamide) hydrogels in water/alcohol mixtures with PC-SAFT, Macromolecules. 45 (16) (2012) 6686–6696, https://doi.org/10.1021/ma300683k.

[25] M.C. Arndt, G. Sadowski, Thermodynamic model for polyelectrolyte hydrogels, J. Phys. Chem. B 118 (35) (2014) 10534–10542, https://doi.org/10.1021/jp501798x.

[26] E.J. García, D. Bhandary, M.T. Horsch, H. Hasse, A molecular dynamics simulation scenario for studying solvent-mediated interactions of polymers and application to Thermoresponse of poly(N-Isopropylacrylamide) in water, J. Mol. Liq. 268 (2018) 294–302, https://doi.org/10.1016/j.molliq.2018.07.025.

[27] I. Alenichev, Z. Sedláková, M. Ilavský, Swelling and mechanical behavior of charged poly(N-Isopropylmethacrylamide) and Poly(N-Isopropylacrylamide) networks in water/ethanol mixtures. Cononsolvency effect, Polym. Bull. 58 (1) (2007) 191–199, https://doi.org/10.1007/s00289-006-0586-3.

[28] G.A. Komarova, E.Y. Kozhunova, I.I. Potemkin, Behavior of PNIPAM microgels in different organic solvents, Molecules. 27 (23) (2022) 8549, https://doi.org/10.3390/molecules27238549.

[29] J.-P. Chen, T.-H. Cheng, Thermo-responsive chitosan-graft-poly(N-Isopropylacrylamide) injectable hydrogel for cultivation of chondrocytes and meniscus cells, Macromol. Biosci. 6 (12) (2006) 1026–1039, https://doi.org/10.1002/mabi.200600142.

[30] L. Zhao, L. Niu, H. Liang, H. Tan, C. Liu, Zhu, F. pH and glucose dual-responsive injectable hydrogels with insulin and fibroblasts as bioactive dressings for diabetic wound healing, ACS Appl. Mater. Interfaces 9 (43) (2017) 37563–37574, https://doi.org/10.1021/acsami.7b09395.

[31] M. Faulde, J. Tonn, A. Jupke, Microgels for the intensification of liquid-liquid extraction processes – Feasibility and advantages, Chem. Eng. Technol. 43 (1) (2020) 137–142, https://doi.org/10.1002/ceat.201900407.

[32] I. Tokarev, V. Gopishetty, J. Zhou, M. Pita, M. Motornov, E. Katz, S. Minko, Stimuli-responsive hydrogel membranes coupled with biocatalytic processes, ACS Appl. Mater. Interfaces 1 (3) (2009) 532–536, https://doi.org/10.1021/am800251a.

[33] K. Miyakawa, F. Sakamoto, R. Yoshida, E. Kokufuta, T. Yamaguchi, Chemical waves in self-oscillating gels, Phys. Rev. E 62 (1) (2000) 793–798, https://doi.org/10.1103/PhysRevE.62.793.

[34] A. Paikar, A.I. Novichkov, A.I. Hanopolskyi, V.A. Smaliak, X. Sui, N. Kampf, E. V. Skorb, s.n. semenov, Spatiotemporal regulation of hydrogel actuators by autocatalytic reaction networks, Adv. Mater. 34 (13) (2022) 2106816, https://doi.org/10.1002/adma.202106816.

[35] D. A. Ossipov, X. Yang, O. Varghese, S. Kootala, J Hilborn, Modular Approach to Functional Hyaluronic Acid Hydrogels Using Orthogonal Chemical Reactions, Chem. Commun. 46 (44) (2010) 8368–8370, https://doi.org/10.1039/C0CC03055D.

[36] H. Wang, S.C. Heilshorn, Adaptable hydrogel networks with reversible linkages for tissue engineering, Adv. Mater. 27 (25) (2015) 3717–3736, https://doi.org/10.1002/adma.201501558.

[37] P.J. Flory, Principles of Polymer Chemistry, Cornell University Press, 1953. ISBN: 978-0-8014-0134-3.

[38] P.J. Flory, J. Rehner, Statistical mechanics of cross-linked polymer networks II. Swelling, J. Chem. Phys. 11 (11) (1943) 521–526, https://doi.org/10.1063/1.1723792.

[39] B. Sung, Mathematical modeling of a temperature-sensitive and tissue-mimicking gel matrix: solving the Flory–Huggins equation for an elastic ternary mixture system, Math. Methods Appl. Sci. 43 (18) (2020) 10637–10645, https://doi.org/10.1002/mma.6855.

[40] T. Lindvig, M.L. Michelsen, G.M. Kontogeorgis, A Flory–Huggins Model Based on the Hansen Solubility Parameters, Fluid. Phase Equilib. 203 (1–2) (2002) 247–260, https://doi.org/10.1016/S0378-3812(02)00184-X.

[41] E. Favre, Q.T. Nguyen, R. Clement, J. Neel, Application of Flory-Huggins theory to ternary polymer-solvents equilibria: a case study, Eur. Polym. J. 32 (3) (1996) 303–309, https://doi.org/10.1016/0014-3057(95)00146-8.

[42] E.J. García, H. Hasse, Studying equilibria of polymers in solution by direct molecular dynamics simulations: poly(N-Isopropylacrylamide) in Water as a Test Case, Eur. Phys. J. Spec. Top. 227 (14) (2019) 1547–1558, https://doi.org/10.1140/epjst/e2018-800171-y.

[43] K. Poschlad, S. Enders, Thermodynamics of aqueous solutions containing poly (N-Isopropylacrylamide) and Vitamin C, Fluid. Phase Equilib. 302 (1–2) (2011) 153–160, https://doi.org/10.1016/j.fluid.2010.08.025.

[44] A. Klamt, Conductor-like screening model for real solvents: a new approach to the quantitative calculation of solvation phenomena, J. Phys. Chem. 99 (7) (1995) 2224–2235, https://doi.org/10.1021/j100007a062.

[45] A. Klamt, V. Jonas, T. Bürger, J.C.W. Lohrenz, Refinement and parametrization of COSMO-RS, J. Phys. Chem. A 102 (26) (1998) 5074–5085, https://doi.org/10.1021/jp980017s.

[46] F. Eckert, A. Klamt, Fast solvent screening via quantum chemistry: COSMO-RS Approach, AIChe J. 48 (2) (2002) 369–385, https://doi.org/10.1002/aic.690480220.

[47] P.B. Staudt, R.L. Simões, L. Jacques, N.S.M. Cardozo, R.D.P. Soares, Predicting phase equilibrium for polymer solutions using COSMO-SAC, Fluid. Phase Equilib. 472 (2018) 75–84, https://doi.org/10.1016/j.fluid.2018.05.003.

[48] C. Loschen, A. Klamt, Prediction of solubilities and partition coefficients in polymers using COSMO-RS, Ind. Eng. Chem. Res. 53 (28) (2014) 11478–11487, https://doi.org/10.1021/ie501669z.

[49] H.S. Elbro, A. Fredenslund, P. Rasmussen, A new simple equation for the prediction of solvent activities in polymer solutions, Macromolecules. 23 (21) (1990) 4707–4714, https://doi.org/10.1021/ma00223a031.

[50] COSMOtherm, Release 19; © 2019; BIOVIA Dassault Systèmes.

[51] COSMOconf, Release 21; © 2021; BIOVIA Dassault Systèmes.

[52] V. Turbomole, 7.3, a Development of university of Karlsruhe and Forschungszentrum Karlsruhe GmbH, 1989–2007, TURBOMOLE GmbH Since 2010 (2007).

[53] R. Putnam, R. Taylor, A. Klamt, F. Eckert, M. Schiller, Prediction of infinite dilution activity coefficients using COSMO-RS, Ind. Eng. Chem. Res. 42 (15) (2003) 3635–3641, https://doi.org/10.1021/ie020974v.

[54] T. Gerlach, S. Müller, A.G. de Castilla, I. Smirnova, An Open Source COSMO-RS Implementation and parameterization supporting the efficient implementation of multiple segment descriptors, Fluid. Phase Equilib. 560 (2022) 113472, https://doi.org/10.1016/j.fluid.2022.113472.

[55] T. Brouwer, B. Schuur, Model performances evaluated for infinite dilution activity coefficients prediction at 298.15 K, Ind. Eng. Chem. Res. 58 (20) (2019) 8903–8914, https://doi.org/10.1021/acs.iecr.9b00727.